\documentclass[aps,prb, twocolumn, superscriptaddress, longbibliography]{revtex4-2}

\usepackage{times}
\usepackage{amsmath}
\usepackage{amssymb}
\usepackage{graphicx}
\usepackage[caption=false, position=top, singlelinecheck=off, justification=raggedright]{subfig}
\usepackage{color}
\usepackage{pst-node}
\usepackage[unicode]{hyperref}
\hypersetup{unicode=true, colorlinks=true, linkcolor=blue, citecolor=blue}
\usepackage{float}
\usepackage[normalem]{ulem}
\usepackage{multirow}
\usepackage{hhline}

\usepackage{mathrsfs,braket}
\usepackage{amssymb, amsbsy, amsmath, latexsym, dsfont, array, layout, graphicx, mathrsfs, color, ulem, bm}

\begin{document}
\title{Spreading dynamics in the Hatano-Nelson model with disorder}
\author{Jinyuan Shang}
\affiliation{Beijing National Laboratory for Condensed Matter Physics, Institute of Physics, Chinese Academy of Sciences, Beijing 100190, China}
\affiliation{School of Physical Sciences, University of Chinese Academy of Sciences, Beijing 100049, China}
\author{Haiping Hu}
\email{hhu@iphy.ac.cn}
\affiliation{Beijing National Laboratory for Condensed Matter Physics, Institute of Physics, Chinese Academy of Sciences, Beijing 100190, China}
\affiliation{School of Physical Sciences, University of Chinese Academy of Sciences, Beijing 100049, China}
\begin{abstract}
The non-Hermitian skin effect is the accumulation of eigenstates at the boundaries, reflecting the system’s nonreciprocity. Introducing disorder leads to a competition between the skin effect and Anderson localization, giving rise to the skin-Anderson transition. Here, we investigate wave packet spreading in the disordered Hatano-Nelson model and uncover distinct dynamical behaviors across different regimes. In the clean limit, transport is unidirectionally ballistic ($\Delta x \sim t$) due to nonreciprocity. For weak disorder, where skin and Anderson-localized modes coexist, transport transitions from ballistic at early times to superdiffusive ($\Delta x \sim t^{2/3}$) at long times. In the deeply Anderson-localized regime, initial diffusion ($\Delta x \sim t^{1/2}$) eventually gives way to superdiffusive spreading. We examine how these scaling behaviors emerge from the system’s spectral properties and eigenstate localization behaviors. Our work unveils the rich dynamics driven by nonreciprocity and disorder in non-Hermitian systems.
\end{abstract}
\maketitle
\section{Introduction}
Non-Hermitian systems exhibit distinct spectral features that sharply contrast with their Hermitian counterparts. A prominent example is the non-Hermitian skin effect (NHSE) \cite{nhse1,nhse2,nhse3,nhse4,nhse5,nhse6,exp1,exp2,exp3,exp4,exp5,exp6,exp7}, accompanied by extreme sensitivity of energy spectra to boundary conditions. In disordered systems, the interplay between non-Hermiticity and randomness leads to intriguing localization effects \cite{nhat1,nhat2,nhat3,nhat4,nhat5,me1,me2,me3,me4,me5,me6,me7,me8,me9,wz_disorder,impurity1,impurity2,impurity3,impurity4,impurity5,impurity6,impurity7,konghao_hu}. One counterintuitive phenomenon is disorder-induced wave propagation \cite{photonic_exp,wz_jump,hh_jump,jumpd1,jumpd2,jumpd3,jumpd4} even when all eigenstates are Anderson-localized modes (ALMs). This apparent inconsistency between Anderson and dynamical localization stems from the jumpy nature of dynamics enabled by complex eigenenergies. Over time evolution, eigenstates with larger imaginary parts of eigenenergies dominate, causing abrupt changes in wave profiles, as observed in photonic lattices with random dissipation \cite{photonic_exp}. A quantitative framework for wave propagation in disordered non-Hermitian systems has recently been developed, revealing a universal relation between the spreading exponent and the imaginary density of states (iDOS) at the band tail \cite{wz_jump,hh_jump}.

The disorder-induced dynamical delocalization marks a fundamental difference between unitary and nonunitary time evolution. In non-Hermitian settings, spreading dynamics is governed by both spectral properties and eigenstate localization. Previous studies \cite{photonic_exp,wz_jump,hh_jump} on wave propagation in disordered non-Hermitian systems have mainly focused on reciprocal cases where the NHSE is absent. In general, adding disorder to a system with the NHSE triggers a gradual skin-Anderson transition \cite{konghao_hu}, as skin modes exhibit resilience against Anderson localization. Below a critical disorder strength, skin modes coexist with the ALMs; beyond this threshold, all eigenmodes become ALMs, and the system enters an Anderson insulator phase. A natural and important question is how wave propagation behaves throughout this transition and how it depends on the underlying spectral properties and eigenstate localization.

In this paper, we investigate wave spreading in the disordered Hatano-Nelson model with nonreciprocity. In the clean limit, this model serves as a protypical example of the NHSE. Using the Lyapunov exponent (LE), we pinpoint the skin-Anderson transition and uncover rich dynamical behaviors across different phases and time scales. i) In the clean case, propagation is unidirectional and ballistic ($\Delta x \sim t$). ii) With weak disorder, ballistic transport persists at first but transitions to superdiffusive transport ($\Delta x \sim t^{2/3}$) at long times. iii) In the strong disorder regime, where the system becomes an Anderson insulator, transport is diffusive ($\Delta x\sim t^{1/2}$) at short times and superdiffusive at long times. We trace these distinct scaling behaviors to the configurations of eigenmodes and the iDOS. Unlike the reciprocal case, our numerics show that directional preference in transport persists across all phases due to nonreciprocity.

The rest of the paper is organized as follows. In Section \ref{secii}, we introduce the Hatano-Nelson model and study its wave spreading, deriving the ballistic transport induced by nonreciprocity. Section \ref{seciii} investigates the skin-Anderson transition, using the transfer-matrix method to determine the critical disorder strength. In Section \ref{seciv}, we examine wave propagation in the disordered case, analyzing distinct scaling behaviors for weak and strong disorder in Sections \ref{seciva} and \ref{secivb}, respectively. Finally, in Section \ref{secv}, we summarize our findings and briefly discuss the experimental implications.

\section{Wave spreading without disorder}\label{secii}
We consider the nonreciprocal Hatano-Nelson model \cite{nhat1} with the Hamiltonian:
\begin{eqnarray}\label{hnmodel}
H_{HN} = \sum_x J e^{g} \ket{x+1} \bra{x} + J e^{-g} \ket{x} \bra{x+1}.
\end{eqnarray}
Here, $g$ characterizes the nonreciprocity, and we set $J=1$ as the energy unit. A nonzero $g$ induces the NHSE, where all eigenstates accumulate at the left (right) boundary for $g<0$ ($g>0$) under open boundary conditions (OBC). The energy spectrum is also highly sensitive to boundary conditions. Under OBC, it takes real values, $E\in[-2J,2J]$, whereas for periodic boundary conditions (PBC), it forms an oval \cite{hu_graph} in the complex plane:
\begin{eqnarray}
E_k = \epsilon_k + i \eta_k = 2J\cosh g \cos k - i 2J \sinh g \sin k,
\end{eqnarray}
with $k\in[0,2\pi]$. Here, $\epsilon_k$ and $\eta_k$ denote the real and imaginary parts of the spectrum, respectively.

\begin{figure}[!t]
\centering
\includegraphics[width=3.33in]{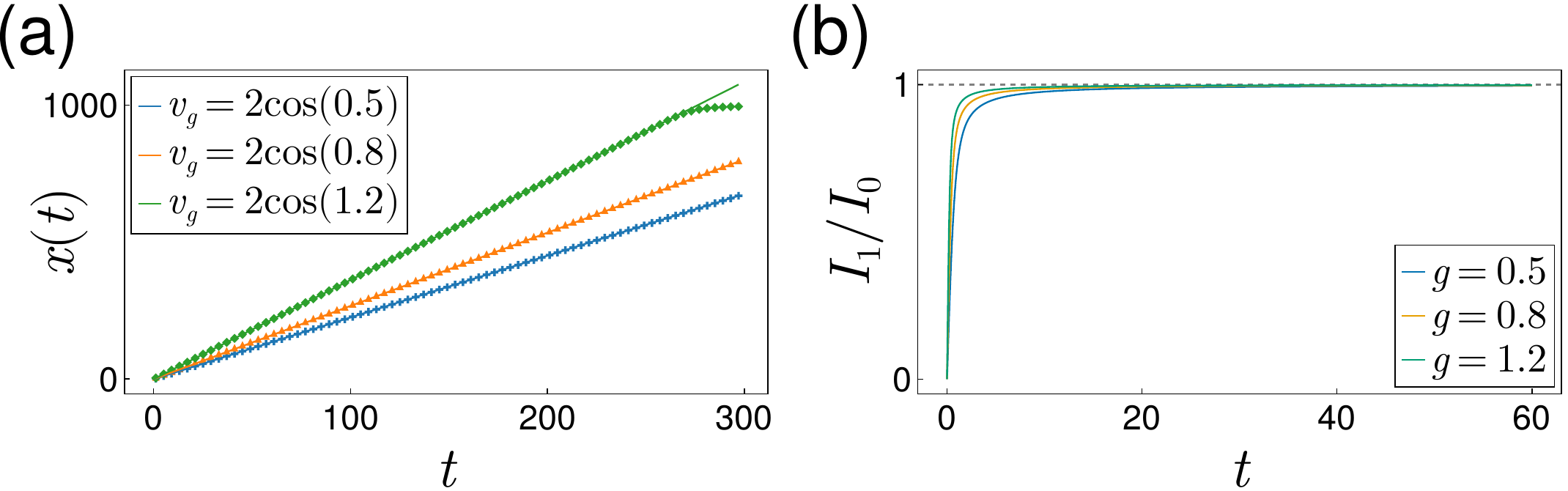}
\caption{Wave spreading in the Hatano-Nelson model [Eq. (\ref{hnmodel})]. (a) The center of mass with respect to time $t$ for $g=0.5$ (blue), $g=0.8$ (orange), and $g=1.2$ (green). The slope, representing the spreading velocity, is well approximated by $v_g=2\cosh g$. The initial packet is prepared at $x_0=0$. (b) Rapid convergence of the ratio of modified Bessel functions of the first kind [See Eq. (\ref{com})] as a function of $t$. }\label{fig1}
\end{figure}
We study wave spreading in a lattice of length $L$, starting with an initial wave packet prepared at the center, $|\psi_0\rangle=|x_0=0\rangle$. The wave packet evolves under the Hamiltonian as $|\psi(t)\rangle=e^{-i H t}|\psi_0\rangle$. Although spectral properties depends on the boundary conditions, the wave dynamics remain unaffected as long as the evolved wave packet has not reached the boundary. Due to the non-Hermitian nature of the Hamiltonian, the evolution is nonunitary, requiring a normalization of $|\psi(t)\rangle$ at each time step. Physically, this corresponds to disregarding non-detection events, as widely used in contexts such as light propagation in photonic lattices \cite{photonic_exp} and discrete-time quantum walks \cite{qw1,qw2,qw3}. At time $t$, the expected center of mass of the wave packet is given by
\begin{eqnarray}\label{com}
x(t) = \frac{\langle\psi(t)|\hat{x}|\psi(t)\rangle}{\langle\psi(t)|\psi(t)\rangle}.
\end{eqnarray}
A numerical simulation of the wave-packet’s evolution is shown in Fig. \ref{fig1}(a). It is clear the transport is ballistic and unidirectional to the right for $g>0$. The group velocity, governed by the nonreciprocity, is well approximated by $v_g=2\cosh g$, indicating that the wave packet moves faster for larger nonreciprocity. Notably, for large $g$, the wavefront distorts at later times due to the finite lattice size, as the wave packet reaches the boundary.

To understand the unidirectionality and ballistic nature of the transport, we solve the wave propagation exactly. The time-evolved wave function can be expanded in the momentum basis as
\begin{eqnarray}
|\psi(t)\rangle = \sum_k \frac{1}{\sqrt{L}} e^{-i\epsilon_k t+\eta_k t} |k\rangle.
\end{eqnarray}
A direct calculation shows that, in the thermodynamic limit, the center of mass evolves as
\begin{eqnarray}
x(t)&=&\frac{\int dk~e^{2\eta_k t}(\partial_k\epsilon_k) t}{\int dk~e^{2\eta_k t}}\notag\\
&=& 2J\cosh g \frac{I_1(4Jt\sinh g)}{I_0(4Jt \sinh g)}t,\label{com}
\end{eqnarray}
where $I_n(.)$ is the modified Bessel function of the first kind \cite{bessel}. For large $t$, the asymptotic expansions of $I_{1,0}(z)$ are
\begin{align}
I_0(z) &\approx \frac{e^z}{\sqrt{2\pi z}} \left(1+\frac{1}{8z}\right), \quad z\to+\infty, \\
I_1(z) &\approx \frac{e^z}{\sqrt{2\pi z}} \left(1-\frac{3}{8z}\right), \quad z\to+\infty.
\end{align}
Figure \ref{fig1}(b) shows $I_1/I_0$ as a function of $t$. It converges rapidly to unity, meaning that the long time evolution is not necessary to observe the ballistic spreading. The transport velocity $v_g$ is given by:
\begin{align}
x(t) &\approx v_gt, \quad v_g = 2 |J|\mathrm{sgn}(g) \cosh g.
\end{align}
The velocity’s sign is given by the sign of $g$ due to the parity of $I_n(z)$, consistent with the intuitive expectation of skin localization.

\section{Skin-Anderson transition}\label{seciii}
We now introduce onsite disorder into the system. The Hamiltonian is
\begin{eqnarray}\label{model2}
H = H_{HN} + \sum_x w_x |x\rangle\langle x|,
\end{eqnarray}
where $w_x$ represents the disorder strength at site $x$. We focus on a simple disorder type—random gain/loss—where $\textrm{Re}[w_x]=0$ and $\textrm{Im}w_x$ is drawn from a uniform distribution over $[-\frac{W}{2},\frac{W}{2}]$, with $W$ denoting the disorder strength. Our analysis can be extended to other disorder types. An immediate consequence of disorder is spectral broadening. In clean systems, the energy spectrum forms arcs or loops under OBC and PBC, while disorder spreads it into finite regions in the complex energy plane. In the thermodynamic limit, the spectral density follows specific distributions which can be obtained through the generalized Thouless relation \cite{konghao_hu,gtr}. 

The disorder drives Anderson localization in the bulk and competes with the NHSE. As the disorder strength increases, skin modes gradually transition into ALMs. This contrasts with the reciprocal case, where all eigenstates undergo Anderson localization at any finite disorder strength. We determine the system’s phase diagram using the transfer-matrix method. Let us consider the eigenvalue equation $H|\phi\rangle=E|\phi\rangle$ with eigenstate $|\phi\rangle=(\phi_1,\phi_2,\cdots,\phi_L)^T$ and eigenenergy $E$. In the coordinate basis, we have
\begin{eqnarray}
Je^{-g}\phi_{x+1}+Je^{g}\phi_{x-1}=(E-w_x)\phi_x.
\end{eqnarray}
 It can be recast into the form $(\phi_{x+1},\phi_x)^T=T_x(\phi_x,\phi_{x-1})^T$, with the one-step transfer matrix taking
\begin{eqnarray}
T_x(E,g)=\left(\begin{array}{cc}
 (E-w_x)J^{-1}e^{g} & -e^{2g} \\
1 & 0\\
\end{array}\right).
\end{eqnarray}
The full transfer matrix, which shifts the left most site to the right most, is given by
\begin{eqnarray}
T(E,g)=\prod_{x=1}^L T_x(E,g).
\end{eqnarray}
It depends on both $E$ and the nonreciprocity $g$. The Lyapunov exponent (LE) is defined as
\begin{eqnarray}\label{le}
\gamma(E,g)=\lim_{L\rightarrow\infty}\frac{1}{L}\ln || T(E,g)||,
\end{eqnarray}
with $||.||$ the matrix norm. Physically, the LE gives the inverse localization length of the eigenstate. For our model (\ref{model2}), an important property is that the LE for different $g$ values are linked by a similarity transformation. Consider the $2\times 2$ matrix $M_x(g)=diag[e^{-x g},e^{-(x-1)g}]$, which relates the one-step transfer matrices for different $g$ as: $T_{x}(E,g_2)=M_{x+1}^{-1}(g_2-g_1)T_x(E,g_1)M_x(g_2-g_1)$. It follows that their LEs satisfy
\begin{eqnarray}\label{rel1}
\gamma(E,g_2)=\gamma(E,g_1)+g_2-g_1.
\end{eqnarray}
This relation implies that the LE for the nonreciprocal case can be obtained from the reciprocal limit $g=0$. 
\begin{figure}[!t]
\centering
\includegraphics[width=3.33in]{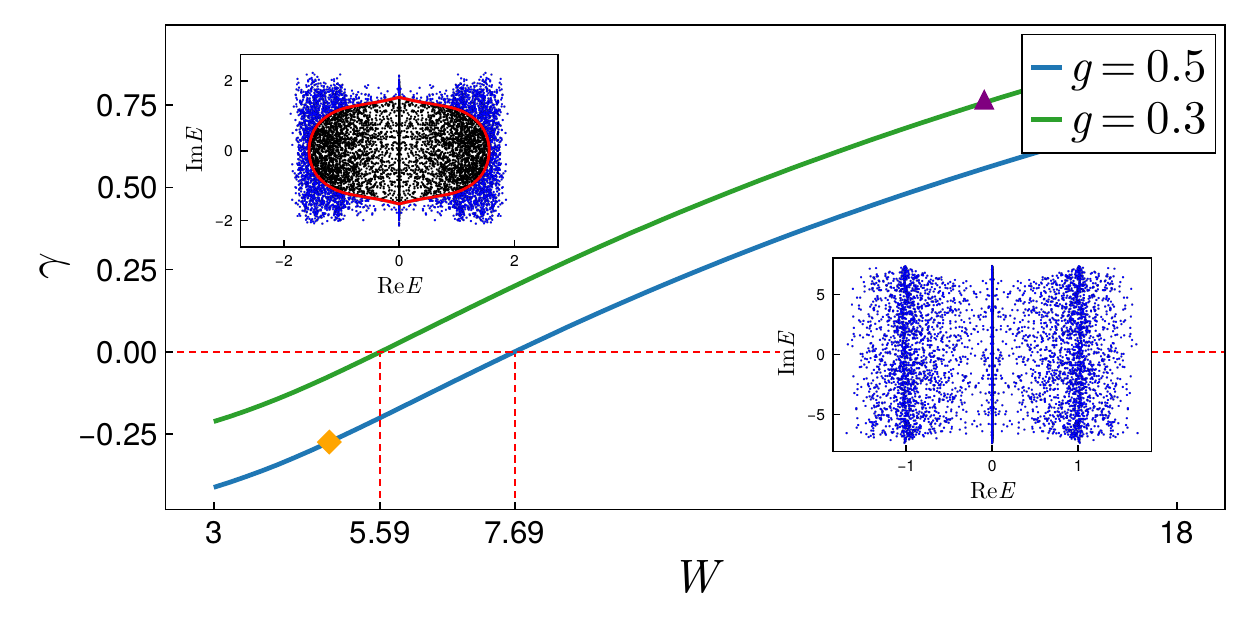}
\caption{Skin-Anderson transition in the disordered Hatano-Nelson model (\ref{model2}). The plot shows the Lyapunov exponent (LE) at $E=0$ versus disorder strength $W$ for $g=0.3$ (green) and $g=0.5$ (blue). The intersection with the red horizontal line marks the threshold disorder strength $W_c$, beyond which the system enters into the Anderson insulator phase. (Insets) Representative energy spectra for the chosen parameters (marked by orange diamond and purple triangle) under open boundary conditions with system size $L=2000$. The skin modes and Anderson localized modes are shown in black and blue, respectively, with the mobility edge highlighted in red.}\label{fig2}
\end{figure}

The skin-Anderson transition can be tracked through the motion of the mobility edge, which separates skin modes from ALMs. The mobility edge is formally determined by the condition
\begin{eqnarray}
\gamma(E,g)=0.
\end{eqnarray} 
Figure \ref{fig2} presents typical OBC energy spectra  for model (\ref{model2}). Notably, they are symmetric with respect to both the real and imaginary axes. The transition begins in the outer spectral region and progressively extends inward as disorder strength increases, accompanied by the shrinking of the mobility edge contour. When the mobility edge vanishes at $E=0$ at a critical disorder strength $W_c$, all skin modes transform into ALMs, marking the onset of the Anderson insulator phase. In Fig. \ref{fig2}, we plot the LE at $E=0$ against the disorder strength for $g=0.3$ and $g=0.5$. As expected, the LEs for different $g$ values differ by a constant, as governed by Eq. (\ref{rel1}). The threshold values are found to be $W_c\approx5.59$ and $W_c\approx 7.69$ for $g=0.3$ and $g=0.5$, respectively. 

\section{Spreading dynamics in the presence of disorder}\label{seciv}
In this section, we examine wave propagation in the disordered Hatano-Nelson model (\ref{model2}) and analyze its scaling behaviors. We focus on two regimes: the weak disorder case, where skin modes coexist with ALMs, and the strong disorder case, where the system falls into the Anderson insulator phase. To characterize the transport, we consider the center of mass $x(t)$, defined in Eq. (\ref{com}), and the absolute deviation from the initial position, given by
\begin{eqnarray}
\Delta x(t)=\sqrt{\sum_x(x-x_0)^2|\psi_x(t)|^2},
\end{eqnarray} 
where $\psi_x(t)$ is the $x$-component of the normalized wave function. Since these quantities depend on specific disorder realizations, we take their ensemble average, $\overline{x(t)}$ and $\overline{\Delta x(t)}$, over disorder samples. The center of mass  quantifies the directional bias of wave spreading, while the behavior of the deviation determines the system’s spreading exponent (dynamical critical exponent).

\subsection{Weak disorder case}\label{seciva}
\begin{figure}[!t]
\centering
\includegraphics[width=3.33in]{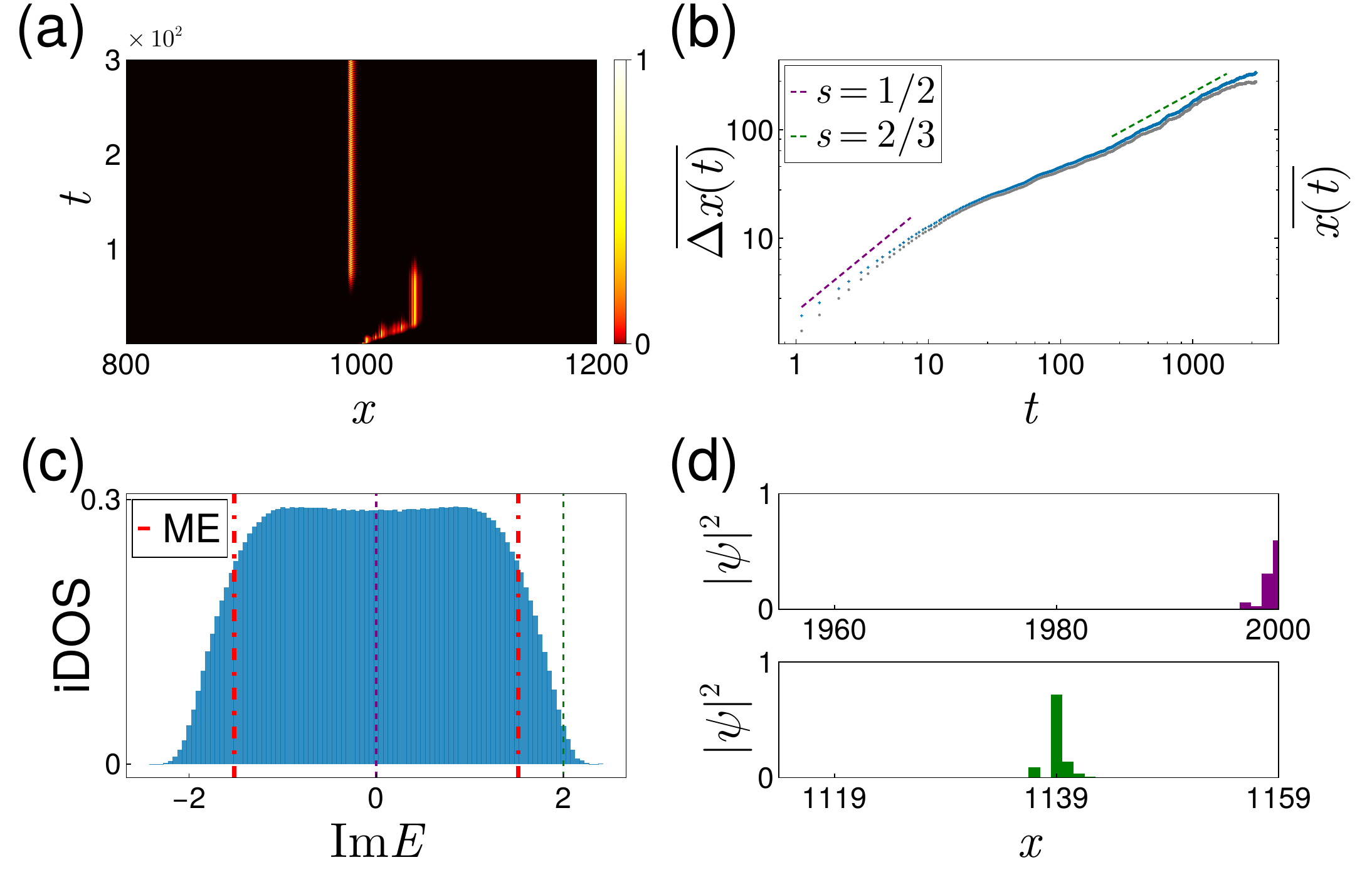}
\caption{Spreading dynamics in the weak disorder regime. (a) Space-time evolution of a wave packet initially prepared at the center of the lattice for a single disorder realization. (b) Ensemble averaged center of mass $\overline{x(t)}$ (gray) and absolute deviation $\overline{\Delta x(t)}$ (blue) as a function of time $t$. $500$ disorder realizations are performed. Dashed lines indicate distinct dynamical scalings at early and late stages. (c) iDOS obtained from the generalized Thouless relation [see Eqs. (\ref{idos1})(\ref{idos2})], with the mobility edge marked by red dashed lines. (d) Representative spatial profiles of the eigenstates (corresponding to the purple and green lines in (c)) at the band center (top panel) and band edge (low panel). The parameters are $W=4.8$, $g=0.5$. The system size is $L=2000$ in (a)(b)(d).}\label{fig3}
\end{figure}

For weak disorder, the dynamics are influenced by both skin modes and ALMs. Figure \ref{fig3}(a) shows the time evolution of a wave packet for a single disorder realization. Initially, the wave packet propagates ballistically and unidirectionally before exhibiting jumpy dynamics. In Fig. \ref{fig3}(b), we plot the ensemble-averaged center of mass $\overline{x(t)}$ (grey line) and absolute deviation $\overline{\Delta x(t)}$ (blue line) as functions of time $t$ (in log scale). Due to nonreciprocity, $\overline{x(t)}\neq 0$. This is in contrast to the reciprocal case where no directional bias appears. The evolution of $\overline{\Delta x(t)}$ shows two distinct stages: ballistic spreading at short times, followed by superdiffusive transport with$\overline{\Delta x(t)}\sim t^{2/3}$ at longer times.

To explain these scaling behaviors, we consider the time-evolved wave function:
\begin{eqnarray}\label{te}
|\psi(t)\rangle=\sum_n a_n e^{-i\epsilon_n t} e^{\eta_n t}|\phi^{(n)}\rangle,
\end{eqnarray}
where $a_n$ represents the overlap of the initial wave packet with the $n$-th eigenstate $|\phi^{(n)}\rangle$ of the Hamiltonian (\ref{model2}). The real part of the eigenenergy $\epsilon_n$ contributes a dynamical phase, while its imaginary part $\eta_n$ modulates the amplitude. Note that the system hosts both skin modes and ALMs separated by the mobility edge. Figure \ref{fig3}(d) plots the spatial profiles of two representative eigenstates near the center and the band edge, which correspond to skin modes and ALMs, respectively. As discussed in Sec. \ref{secii}, skin modes induce ballistic and unidirectional transport. Let $\eta_{\text{max}}$ and $\eta_{\text{ME}}$ denote the location of the band edge and mobility edge. From Eq. (\ref{te}), the ballistic transport should persists up to a time scale of 
\begin{eqnarray}
t_{\mathrm{ballistic}}\sim\frac{1}{\eta_{\mathrm{max}}-\eta_{\mathrm{ME}}},
\end{eqnarray}
after which the ALMs near the band edge take over.

The spreading driven by the ALMs are of jumpy nature, with the spreading exponent determined by the iDOS near the band tail \cite{wz_jump,hh_jump}. Formally, the iDOS can be extracted using the generalized Thouless relation \cite{gtr,konghao_hu}:
\begin{eqnarray}\label{idos1}
\rho(E)=\frac{1}{2\pi}\nabla^2\gamma(E),
\end{eqnarray}
where $\rho(E)$ is the spectral density and $\gamma(E)$ is the LE. Integrating over the real part of the eigenenergies yields the iDOS:
\begin{eqnarray}\label{idos2}
\rho_I(\eta)=\int \rho(\epsilon+i\eta)~d\epsilon.
\end{eqnarray}
This approach avoids diagonalizing large non-Hermitian matrices or averaging over disorder realizations. Figure \ref{fig3}(c) plots the resulting iDOS, which exhibits a trapezoidal shape with a linear drop at the band edge. According to the scaling theory of jumpy dynamics \cite{wz_jump,hh_jump}, the spreading exponent dictated by the band-tail states is $s = 2/3$, matching the long-time numerical results in Fig. \ref{fig3}(b).

\subsection{Strong disorder case}\label{secivb}
In this subsection, we examine the strong disorder regime. Although all eigenstates are bulk-localized in the Anderson insulator phase, wave spreading can still occur due to disorder-induced dynamical delocalization. In Fig. \ref{fig3}(a), we show the time evolution of the wave packet for a single disorder realization, where the jumpy nature of non-Hermitian dynamics is clearly visible. Fig. \ref{fig3}(b) displays the ensemble-averaged center of mass, $\overline{x(t)}$, and absolute deviation $\overline{\Delta x}(t)$ over time. Unlike reciprocal disordered non-Hermitian systems, a directional bias persists here. The transport exhibits two distinct scaling regimes: at short times, $\overline{\Delta x(t)}\sim t^{1/2}$, while at long times, $\overline{\Delta x(t)}\sim t^{2/3}$. In Fig. \ref{fig3}(c), we show the iDOS obtained from the LE. It exhibits a trapezoidal structure—a platform at the band center and linear drop at the edges. The platform is reminiscent of the uniform distribution of the imaginary disorder. Fig. \ref{fig3}(d) further shows typical spatial profiles of eigenstates: those near the band center are more extended (larger localization length), while those at the edges are confined to fewer sites. This suggests that they originate from rare disorder realizations where neighboring onsite potentials take nearly identical values, leading to a linear tail in the iDOS \cite{wz_jump}.
\begin{figure}[!t]
\centering
\includegraphics[width=3.33in]{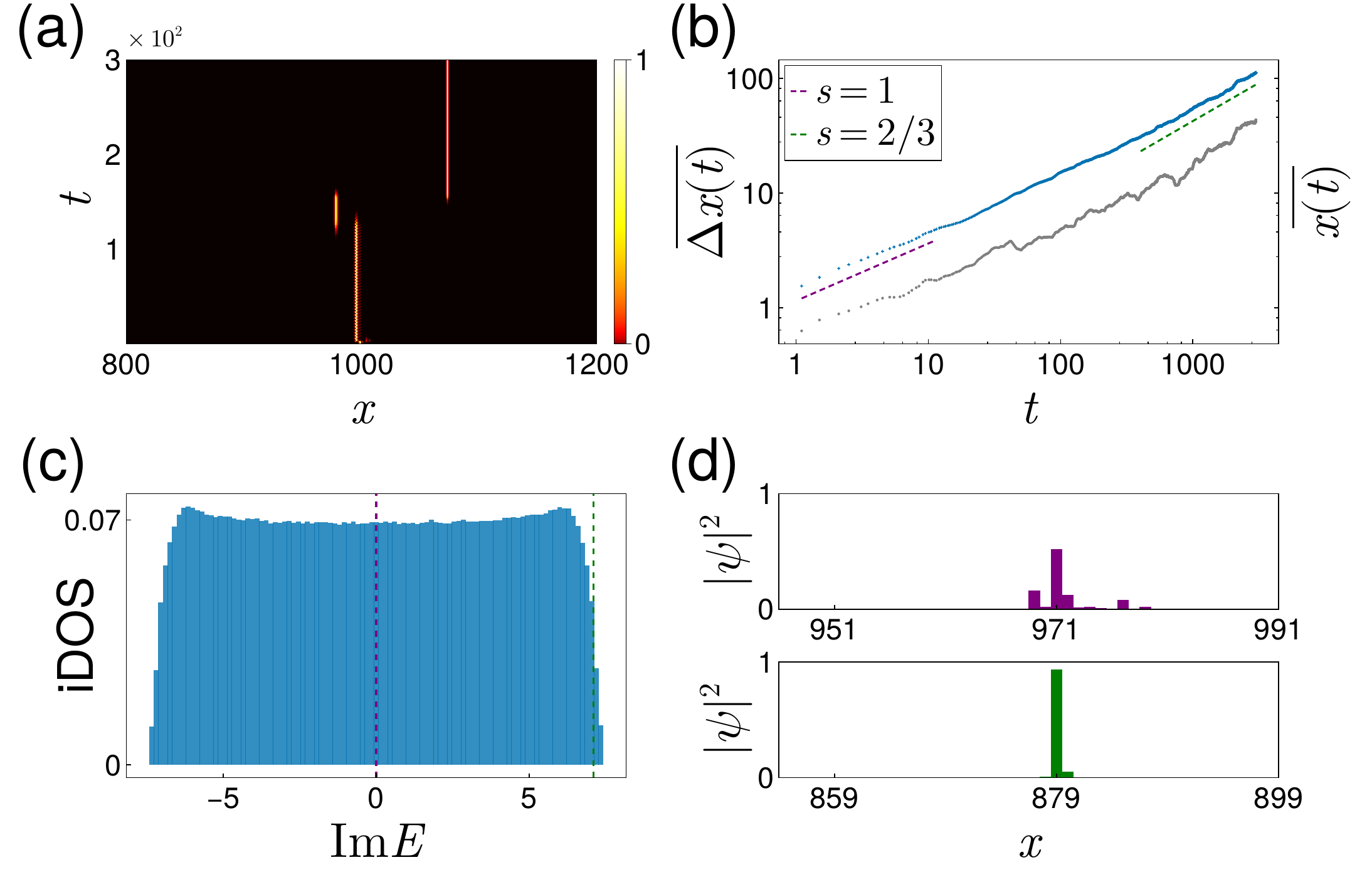}
\caption{Spreading dynamics in the strong disorder regime. (a) Space-time evolution of a wave packet initially prepared at the center of the lattice for a single disorder realization. (b) Ensemble averaged center of mass $\overline{x(t)}$ (gray) and absolute deviation $\overline{\Delta x(t)}$ (blue) as a function of time $t$. $500$ disorder realizations are performed. Dashed lines indicate distinct dynamical scalings at the early and late stages. (c) iDOS obtained from the generalized Thouless relation. (d) Representative spatial profiles of the eigenstates (corresponding to the purple and green lines in (c)) at the band center (top panel) and band edge (low panel). The parameters are $W=15$, $g=0.3$. The system size is $L=2000$ in (a)(b)(d).}\label{fig4}
\end{figure}

The scaling behavior at different time stages can be understood as follows. At short times, wave evolution involves contributions from all eigenstates. Since eigenstates at the band tail arise from rare disorder events, the scaling is governed by the platform structure of the iDOS. The scaling theory of jumpy dynamics predicts a spreading exponent of $s=1/2$, indicating diffusive transport. At long times, eigenstates near the band edge dominate, and the linear drop in the iDOS leads to a spreading exponent of $s=2/3$, corresponding to superdiffusive transport. A similar scaling crossover also occurs in the reciprocal case \cite{wz_jump}. However, in the nonreciprocal case, ALMs exhibit spatially asymmetric decay rates. This introduces a directional bias in the center of mass but does not affect the long-time scaling. Note that in the scaling theory of jumpy dynamics \cite{wz_jump,hh_jump}, all eigenmodes are assumed to have a typical localization length. We argue that nonreciprocity has little impact on the localization properties of eigenstates at the band edge. In our model, the reciprocal and nonreciprocal cases are connected by a similarity transformation, which preserves the energy spectrum under OBC. The localization length of an eigenmode with energy $E$ in the nonreciprocal case satisfies 
\begin{eqnarray}
\frac{1}{\xi(E)}=\frac{1}{\xi_0(E)}-g,
\end{eqnarray}
where $\xi_0(E)$ is the localization length at $g=0$. Since long-term dynamics are governed by eigenstates with small $\xi_0$ at the band edge, we have $\xi(E)\approx \xi_0(E)$.

\section{Conclusion and discussion}\label{secv}
To conclude, we have investigated the disordered Hatano-Nelson model and uncovered rich spreading dynamics across different regimes and time scales. In the clean limit, propagation is unidirectional and ballistic. As disorder increases, ballistic transport persists initially but eventually gives way to superdiffusive behavior at long time. In the strong disorder regime, where only ALMs are present, transport is diffusive at early times before becoming superdiffusive at later stages. These transport behaviors are summarized in Table \ref{tablei}. We analyze the distinct scalings through the lens of the iDOS and the configurations of different eigenmodes. Unlike the reciprocal case, we find that the directional preference always exists in the nonreciprocal system.
\begin{table}[!h]
    \caption{Wave spreading in different regimes and time stages for the disordered Hatano-Nelson model.}
    \label{tab:case}
    \centering
    \begin{tabular}{|c|c|c|c|}
        \hline
        cases & clean case & weak disorder & strong disorder \\
        \hline
        scalings & $x\sim vt$ & $x\sim t \rightarrow t^{2/3}$ & $x\sim t^{1/2}\rightarrow t^{2/3}$\\
        \hline
    \end{tabular}\label{tablei}
\end{table}

Although we focus on the simplest Hatano-Nelson model, our analysis extends to more complex models and other types of disorder. Notably, a complex Thouless relation \cite{konghao_hu,gtr} connects the spectral density to the LEs, enabling efficient evaluation of the iDOS and its band-tail behavior without large-scale numerical diagonalization of disordered non-Hermitian Hamiltonians. Our work highlights the intricate interplay between the NHSE and disorder, with potential extensions to disordered or randomly dissipative open quantum systems governed by Lindbladian master equations \cite{jumpd2}. An intriguing direction is to explore how this interplay influences non-Hermitian dynamics in higher dimensions. On one hand, dimensionality plays a crucial role in Anderson transitions and the scaling relations in jumpy dynamics \cite{wz_jump,hh_jump}; on the other, the NHSE becomes significantly richer in higher dimensions \cite{fc_gdse,wz_amoeba,hu_uniform,hu_nonbloch,zhangkai,zhangkai2}, giving rise to diverse types of skin modes. Experimentally, we expect disorder-induced spreading dynamics in nonreciprocal systems to be observable in platforms such as photonic waveguides \cite{photonic_exp} or quantum walks \cite{qw1,qw2,qw3} with well-controlled disorder and dissipation.

\begin{acknowledgments}
This work is supported by the National Key Research and Development Program of China (Grants No. 2023YFA1406704 and No. 2022YFA1405800) and National Natural Science Foundation of China (Grant No. 12474496).
\end{acknowledgments}

\end{document}